\documentstyle[aas2pp4,psfig]{article}
\newcommand\ns{neutron star}
\newcommand\xr{X-ray}
\newcommand\lmfns{low-magnetic-field neutron star}
\newcommand\ccd{color-color diagram}
\newcommand\mdot{$\dot M$}
\newcommand\psm{power spectrum}
\newcommand\sco{\hbox{Sco\,X-1}}
\newcommand\usec{$\mu$s}
\newcommand\Sz{S$_{\rm Z}$}
\newcommand\nnfbo{\nu_{_{\rm N/FBO}}}

\newcommand\pp{$\pm$}

\def\hide#1{}
\newcommand\psl{power-spectral}

\righthead{Sub-Millisecond QPO in Sco X-1}
\slugcomment{To appear in ApJ Letters, scheduled for the issue of September
20, 1996} 
\begin{document}

\title{\noindent{\normalsize\tt To appear in ApJ Letters, scheduled for the issue of September
20, 1996\vskip 24pt}Discovery of Sub-Millisecond Quasi-Periodic Oscillations\\
in the X-Ray Flux of Scorpius\,X-1}


\author{M. van der Klis}
\affil{Astronomical Institute ``Anton Pannekoek'', University of
Amsterdam \\and Center for High-Energy Astrophysics, Amsterdam}

\author{J.H. Swank, W. Zhang, K. Jahoda}
\affil{Goddard Space Flight Center, NASA}

\author{E.H. Morgan, W.H.G. Lewin}
\affil{Massachusetts Institute of Technology}

\author{B. Vaughan}
\affil{California Institute of Technology}

\and

\author{J. van Paradijs}
\affil{University of Alabama at Huntsville and University of Amsterdam}

\begin{abstract}
We report the discovery, with NASA's Rossi X-ray Timing Explorer (RXTE), of
the first sub-millisecond oscillation found in a celestial X-ray source. The
quasi-periodic oscillations (QPO) come from Sco X-1 and have a frequency of
approximately 1100\,Hz, amplitudes of 0.6--1.2\% (rms) and are relatively
coherent, with Q up to $\sim$10$^2$. The frequency of the QPO increases with
accretion rate, rising from 1050 to 1130\,Hz when the source moves from top to
bottom along the normal branch in the X-ray color-color diagram, and shows a
strong, approximately linear correlation with the frequency of the well-known
6--20\,Hz normal/flaring branch QPO. We also report the discovery of QPO with
a frequency near 800\,Hz that occurs, simultaneously with the 1100\,Hz QPO, in
the upper normal branch. We discuss several possible interpretations, one
involving a millisecond X-ray pulsar whose pulses we see reflected off
accretion-flow inhomogeneities. Finally, we report the discovery of
$\sim$45\,Hz QPO, most prominent in the middle of the normal branch, which
might be magnetospheric beat-frequency QPO.
\end{abstract}

\keywords{stars: individual (Sco X-1) --- stars: neutron --- pulsars: general}

\section{Introduction}

The characteristic time scale for motion of matter near a gravitating object,
the dynamical time scale $\tau_{dyn}$, is $(r^3/GM)^{1/2}$, where $r$ is the
distance to the object and $M$ its mass. For $r=R$, the size of the object,
$\tau_{dyn}$ is relevant to possible spin and vibration periods. For a \ns\
$\tau_{dyn}$ extends down to values of well below 1\,ms.  Sub-millisecond
variability could therefore be produced in an accreting neutron star by
various different mechanisms. In this paper, we present the first conclusive
evidence for sub-millisecond variability in an accreting neutron star,
Sco\,X-1. A preliminary announcement of this work was already made in Van der
Klis et al. (1996).

Sco\,X-1 (Giacconi et al.\ 1962), the brightest persistent \xr\ source in
the sky, is a Z source (Hasinger and Van der Klis 1989), a luminous low-mass
\xr\ binary containing a \lmfns. In the \xr\ \ccd\ it shows the lower two of
the three branches in the canonical ``Z track'', the normal branch (NB) and
the flaring branch (FB) (Middleditch and Priedhorsky 1986, Priedhorsky et al.\
1986, Hertz et al.\ 1992, Dieters and Van der Klis 1996). Its \psm\ is
characterized by the presence of 1--5\% amplitude, 6--20\,Hz normal/flaring
branch QPO (N/FBO). According to the standard interpretation (e.g., Hasinger
et al. 1990, Lamb 1991), when the mass transfer rate \mdot\ to the \ns\
increases, the source moves down along the NB, passes through the ``vertex''
of the two branches, where it reaches the Eddington critical rate, and then
moves up the FB. At the same time the QPO frequency increases from $\sim$6\,Hz
on the NB to $\sim$20\,Hz on the lower FB. Further up the FB the QPO
disappear.  Below, we report three new, previously unknown QPO phenomena in
this source, and describe their dependence on \mdot.

\section{Observations}

Sco\,X-1 was observed three times with the proportional counter array (PCA)
onboard NASA's Rossi X-ray Timing Explorer (RXTE; Bradt, Rothschild and Swank
1993), on 1996 Feb.\,14 from 9:14 to 13:25\,UT, Feb.\,18 4:46--8:40\,UT and
Feb.\,19 10:09--14:56\,UT, hereafter observations 1, 2 and 3,
respectively. Each observation covered three satellite orbits with
$\sim$2500\,s of data, separated by intervals of $\sim$1500\,s due to Earth
occultation and/or South-Atlantic Anomaly passage. The count rate was
$\sim10^5$\,c/s in observations 1 and 2. At the start of orbit 1 of
observation 3 the count rate increased to $\sim$1.45\,10$^5$\,c/s, which
tripped the PCA high-voltage safety switch-off. RXTE was then moved slightly
off \sco\ to $\sim$50\% collimator efficiency; later count rates were between
0.5 and 0.8\,10$^5$\,c/s. The $<$20\,keV background was $\lesssim$90\,c/s and
has been negelected.

\begin{figure}[hbt]
\begin{center}
\begin{tabular}{c}
\psfig{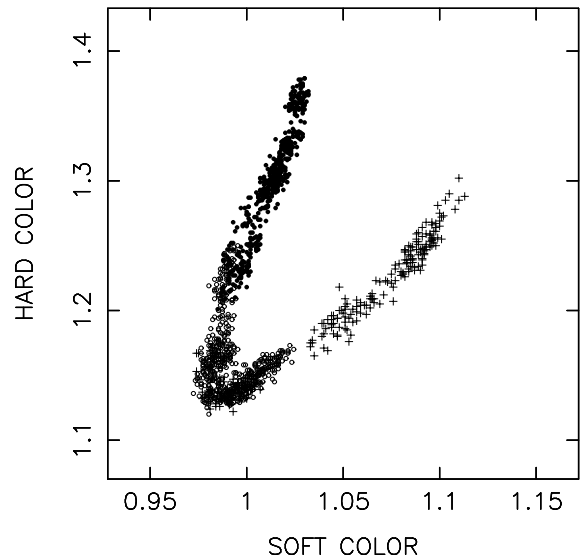}
\end{tabular}
\caption{X-ray color-color diagram. Soft color is the
(3-5)/(1-3)~keV, hard color the (7-20)/(5-7)~keV count-rate ratio.  Each data
point corresponds to 16\,s of data; filled circles, open circles and crosses
correspond to observations 1, 2 and 3, respectively. The statistical
uncertainties are $\sim$0.003 in soft color and $\sim$0.005 in hard color.
The systematic uncertainties in Z track location between
observations (see text) are $\sim$1\%. \label{fig1}}
\end{center}
\end{figure}

During orbits 1 and 3 of each observation, data were collected with a time
resolution of initially 0.5\,ms and later, after discovery of the 1100\,Hz QPO
in observation 1, usually 0.25\,ms. During orbit 2 of each observation, the
time resolution was set to 16\,\usec, and ``double events'' were additionally
recorded at 64\,\usec\ time resolution. Double events, two events detected
within 6\,\usec\ at two different anodes of a PCA detector, are normally
mostly due to charged particles. However, for the very high count rates from
\sco\ they are mostly due to two source photons. In the data analysis
the double events, counted as two, were added to the single-photon data. This
led to a considerable increase in sensitivity to time variability. All
high-time-resolution data were recorded in the 2--20\,keV band. During the
entire run 16-s resolution data were additionally recorded in 129 spectral
channels covering the 2--60\,keV band.

\section{Analysis and Results}

\begin{figure}[htb]
\begin{center}
\begin{tabular}{c}
\psfig{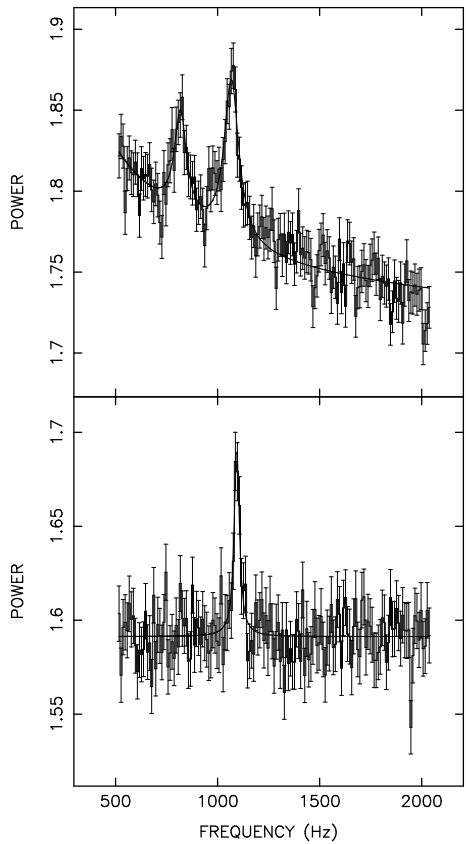}
\end{tabular}
\caption{Power spectra of combined single- and
double-event data (see text) showing simultaneous 800 and 1100\,Hz QPO ({\it
top}; orbit 2 of observation 1), and a narrow 1100\,Hz QPO peak ({\it bottom};
orbit 2 of observation 2). The spectra are Leahy-normalized; the offsets of
the continua from a level of 2.0 and the slope of the continuum in observation
1 are due to instrumental deadtime effects. \label{fig2}}
\end{center}
\end{figure}

We used the 16-s data to construct \xr\ \ccd s. There was a considerable
offset between the Z track seen in observations 1 and 2 and that in
observation 3, which we attribute to energy-dependent detector effects
(collimator reflections and/or rate-dependent gains). After an empirical
correction that assumes the Z track did not move between our observations
(e.g., Dieters and Van der Klis 1996), a typical \sco\ \ccd\ results
(Fig.\,1). During observation 1 the source moved gradually down the NB, during
observation 2 it moved up and down between the lower NB and FB, and in
observation 3 it moved gradually down the FB and into the vertex region. We
assigned numbers \Sz\ to positions on the Z track according to curve length
along the track, where we set the top of our NB to 1, and the vertex to 2.

Approximately 1100-Hz QPO (Fig.\,2) were observed on all occasions that the
source was on the NB or very low on the FB, and the observational set-up was
adequate (full collimator effciency to ensure sufficient sensitivity, and
sufficient time resolution). This was the case in orbit 2 of observation 1 and
in orbits 1 and 2 of observation 2. The QPO were seen independently in single
and in double-event data. Table\,1 lists the results of functional fits of a
constant plus a power law plus one or two Lorentzian peaks to power spectra of
the high time resolution data. All fits were statistically acceptable with
reduced $\chi^2\sim1$. The QPO frequency, $\nu_{1100}$, increased with \mdot,
from 1050\,Hz at \Sz\ = 1.25 to 1130\,Hz at \Sz\ = 2.1. Note that for lack of
time resolution we have no information about the 1100\,Hz QPO in the \Sz\
range 1.4--1.8. The QPO were sometimes quite coherent, with $Q\equiv
\nu/\Delta\nu$ up to $10^2$. Fractional rms amplitudes, corrected for
differential deadtime (Van der Klis 1989) assuming a paralyzable process
(Zhang et al. 1995) with a deadtime of 10\,\usec\ (Zhang 1995) ranged between
0.9 and 1.2\% in observation 1 and 0.6--0.9\% in observation 2. There was no
clear dependence on \mdot\ within each observation.

QPO with a frequency of approximately 800\,Hz were observed simultaneously
with the 1100-Hz QPO (Fig.\,2 {\it (top)}) only when the source was in the
upper NB (\Sz$<$1.36). From studying the short-term \psl\ variations we can
exclude that the two peaks are due to one peak moving in frequency, unless it
moves on time scales shorter than 32\,s. In orbit 2 of observation 1 the QPO
frequency, $\nu_{800}$, generally increased (from $\sim$800 to $\sim$830\,Hz)
when \Sz\ rose from 1.25 to 1.35 ($\nu_{1100}$ increased from 1050 to 1075\,Hz
over the same range). The 800\,Hz QPO were relatively broad, 50 to $>$100\,Hz
and had amplitudes between 0.9 and 1.2\% (rms), decreasing slightly with
\Sz. In orbit 1 of observation 1 (\Sz$<$1.16) the sensitivity was less due to
the lack of double-event data, but there was some evidence (3$\sigma$) for the
presence of a broad peak near 720\,Hz.

A third new QPO feature, located near 45\,Hz, was detected on the NB. These
QPO had a FWHM of typically 15--25\,Hz, an amplitude of $\sim$1\% (rms) and
were most prominent on the middle of the NB, near \Sz\ = 1.5. They were
detected all the way from \Sz\ = 1.1 to 1.9 with little change in frequency.
Fig.\,3 shows the power spectrum of data obtained in orbit 3 of observation 1,
with 6\,Hz QPO (the N/FBO) and 45\,Hz QPO.

\begin{figure}[htb]
\begin{center}
\begin{tabular}{c}
\psfig{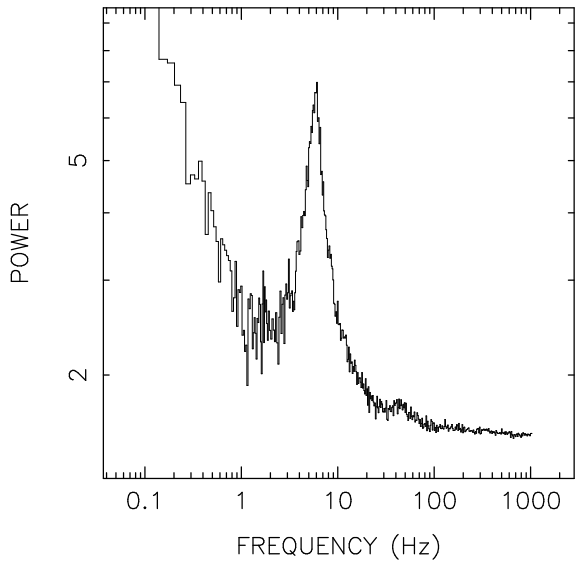}
\end{tabular}
\caption{Power spectrum of 0.5\,ms resolution data 
obtained in orbit 3 of observation 1, showing a strong N/FBO peak centered on
6\,Hz, and a weaker feature near 45\,Hz. The spectrum is Leahy-normalized; the
offset of the continuum from a level of 2.0 is due to deadtime effects.
\label{fig3}}
\end{center}
\end{figure}

The dependencies of $\nu_{1100}$ and the N/FBO frequency $\nnfbo$ on \Sz\ are
very similar. The relation between $\nu_{1100}$ and $\nnfbo$ is plotted in
Fig.\,4. There is a strong correlation. The relation fits a straight line,
with $\nu_{1100} = 7.3\cdot\nnfbo + 1032$; a power law fits as well.

\begin{figure}[htb]
\begin{center}
\begin{tabular}{c}
\psfig{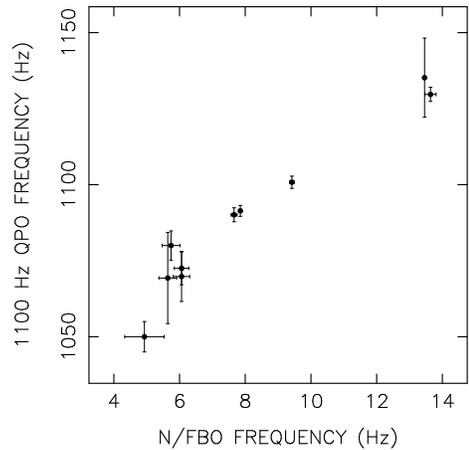}
\end{tabular}
\caption{Relation between the frequencies of the 
1100\,Hz QPO and the N/FBO as derived from data where these two phenomena were
detected simultaneously. A strong correlation is evident.
\label{fig4}}
\end{center}
\end{figure}

\section{Discussion}

The discovery of near-millisecond (Strohmayer, Zhang and Swank 1996, Van
Paradijs et al. 1996, this paper) and submillisecond (Van der Klis et
al. 1996, this paper) oscillations from accreting \lmfns s was expected on the
basis of general arguments (see Introduction). To find and study such
phenomena was one of the primary objectives of the RXTE mission. There is a
range of phenomena that could take place on these time scales, and we shall
not attempt to be exhaustive here.  The fact that both $\nu_{1100}$ and
$\nu_{800}$ increase considerably as a function of \mdot\ suggests an origin
in accretion phenomena rather than \ns\ vibrations (e.g., McDermott, Van Horn
and Hansen, 1988). Therefore, we concentrate on that class of models.

{\it Disk frequency.} The Kepler frequency $\nu_K$ at the inner edge of the accretion
disk, bounded by a small (16--20\,km) magnetosphere is in the correct range to
be identified with either $\nu_{1100}$ or $\nu_{800}$. It would increase with
\mdot, as observed. The \mdot\ dependence of the 800\,Hz QPO, strongest at low
 \mdot\ and disappearing towards higher \mdot\ is reminiscent of the
``horizontal branch QPO'' (HBO) in other Z sources (Van der Klis et al. 1985,
Van der Klis 1995 for a review). These HBO, in the beat-frequency model (Alpar
and Shaham 1985, Lamb et al.\ 1985), are caused by ``clumps'' in the inner
disk, which are most prominent at the lowest \mdot\ levels. The 800\,Hz QPO
(or, alternatively, the 1100\,Hz QPO) could be a direct signal from these
clumps, either by reflection or obscuration of the \xr\ flux from the central
regions, or by interaction with azimuthal structure in the radial flow (see
below).

{\it Beat frequency.} As no HBO have been detected as yet from \sco, an
alternative possibility is that the 800\,Hz QPO are HBO, and their frequency
the beat frequency $\nu_B = n(\nu_K-\nu_s)$ between $\nu_K$ and the \ns\ spin
frequency $\nu_s$, where $n$ is the field symmetry factor.  This would make
the HBO in \sco\ {\it much} faster than those in other Z sources, whose
frequency is 15--55\,Hz, and would require a relatively slow \ns\ spin rate
(not more than a few 100\,Hz). When this Letter was about to be submitted, the
detection was reported in 4U\,1728$-$34 of 363\,Hz oscillations during X-ray
bursts (Strohmayer et al. 1996) which were interpreted as due to the neutron
star spin, and two variable-frequency QPO peaks (one of which the
$\sim$800\,Hz QPO reported by Strohmayer, Zhang and Swank, 1996) whose
separation remained near 363\,Hz. Clearly, this suggests a beat frequency
model interpretation wit $\nu_B$, $\nu_K$ and $\nu_s$ all visible. In the case
of \sco, the difference $\nu_{1100}-\nu_{800}$ is consistent with being
constant near 250\,Hz, but the range in frequencies over which both peaks are
detected simultaneously is too small to draw any strong conclusion from this;
there is no evidence for an oscillation near 250\,Hz.

The new 45\,Hz QPO in \sco\ {\it are} in the range of HBO frequencies seen in
the other Z sources. The dependence of their amplitude on \Sz\ is clearly
different from that of HBO in GX\,5$-$1 and Cyg\,X-2, which are strongest at
the top of the NB (and in the HB), not in the middle of the NB. In GX\,17+2,
60\,Hz QPO have been discovered in the NB which may be a similar phenomenon
(Wijnands et al. 1996, in prep.).  If the 45\,Hz QPO in \sco\ are
beat-frequency QPO, then from the fact that its frequency is approximately
constant we conclude that the 1100\,Hz and the 800\,Hz QPO, whose frequencies
vary by tens of Hz, can not be the disk frequency.

{\it Doppler-shifted pulsar.} The strong correlation we observe between
$\nu_{1100}$ and $\nnfbo$ is not explained by any of the models discussed
above. It could in principle arise because both frequencies vary as a
function of \mdot\ independently of each other, but this seems unlikely. The
$\nnfbo$ vs. \Sz\ relation is known to jump at \Sz=2 (see, e.g., Van der Klis
1995), and $\nu_{1100}$ should then, independently, show a similar jump in its
dependence on \Sz\ to maintain the correlation. It seems more likely that one
frequency is by some mechanism directly derived from the other one.

We briefly explore a mechanism where this is the case.  In the Fortner, Lamb
and Miller (1989) model for N/FBO, part of the accretion takes place by way of
an approximately radial inflow, and blobs (inhomogeneities) in this flow
produce the N/FBO. For a constant-size radial flow, $\nnfbo$ is proportional
to the inflow velocity $v_r$. If the 1100\,Hz QPO are the Doppler-shifted
pulsar signal, which we see reflected off blobs in the flow, then when $v_r$
varies, $\nu_{1100}$ would vary linearly with $\nnfbo$, as observed.  For a
blob moving towards the star along a radius vector that makes an angle $\phi$
with the line of sight, $\nu_{1100}$ would be proportional to
$(1+(1-\cos\phi)v_r/c)\nu_{puls}$, where $\nu_{puls}$ is the pulse frequency.
Radial flow velocities up to at least 0.04$c$, and spin frequencies of
$\sim$500 or $\sim$1000\,Hz, depending on whether one or two magnetic poles
contribute to the signal, are required for this model to work. To get the
relatively coherent QPO signal we observe, the range of angles $\phi$
contributing to the signal should be small. This would be true, for example,
if the magnetic and rotation axes are nearly aligned.  Aligned rotation would
also explain why millisecond pulsars in low-mass \xr\ binaries have been hard
to find. This model would require a spin period $\sim$1\,ms, faster than any
\ns\ spin yet measured, as only one pole would contribute (the other pole
could produce another QPO peak).

{\bf Note.} Further RXTE observations of \sco\ on May 24-25 again show the two
high-frequency QPO peaks, this time with frequencies near 900 and 600\,Hz.
Their separation is significantly ($>$10$\sigma$) larger than in the February
observations. This excludes the beat-frequency interpretation where the peak
separation is predicted to be constant at the neutron star spin frequency.

\acknowledgments

This work was supported in part by the Netherlands Organization for Scientific
Research (NWO) under grant PGS 78-277 and by the Netherlands Foundation for
Research in Astronomy (ASTRON) under grant 781-76-017. WHGL and JVP
acknowledge support from the National Aeronautics and Space Administration.

\begin{deluxetable}{ccccccccc}
\tablecolumns{9}
\footnotesize
\tablewidth{0pt}
\tablecaption{QPO parameters \label{tab1}}
\tablehead{
\colhead{\Sz} & \multicolumn{2}{c}{1100\,Hz} && \multicolumn{2}{c}{800\,Hz} && 
\multicolumn{2}{c}{N/FBO} \\
\cline{2-3} \cline{5-6} \cline{8-9} \\
\colhead{} & \colhead{Frequency} & \colhead{FWHM} && \colhead{Frequency} &
\colhead{FWHM} && \colhead{Frequency} & \colhead{FWHM} \\
\colhead{} & \colhead{(Hz)} & \colhead{(Hz)} && \colhead{(Hz)} & \colhead{(Hz)} &&
\colhead{(Hz)} & \colhead{(Hz)} }
\startdata
\sidehead{Observation 1 orbit 2.}
1.252\pp0.027& 1050.3\pp1.1&               41\pp5&&  803\pp3&            88\pp11&&    4.9\pp0.6&      13\pp3\nl
1.282\pp0.040& 1069.8\pp8.2&             104\pp30&& 794\pp20& 145$_{-50}^{+189}$&&  6.06\pp0.25&   9.3\pp1.1\nl
1.336\pp0.032& 1080.0\pp4.8&              50\pp19&& 833\pp12&             80 fix&&  5.74\pp0.28&  10.7\pp1.1\nl
1.340\pp0.050& 1072.5\pp5.4&              78\pp18&&  829\pp5&            49\pp19&&  6.06\pp0.22&   8.5\pp0.8\nl
1.357\pp0.022&    1069\pp15&   178$_{-65}^{+149}$&& 820\pp12&             80 fix&&  5.64\pp0.27&   9.9\pp1.1\nl
\sidehead{Observation 2 orbit 2.}
1.866\pp0.022& 1091.4\pp1.8&               22\pp5&&      ---&                ---&&  7.85\pp0.04& 4.44\pp0.12\nl
1.929\pp0.050& 1100.8\pp2.0&               18\pp4&&      ---&                ---&&  9.42\pp0.07& 7.24\pp0.23\nl
\sidehead{Observation 2 orbit 1.}
1.966\pp0.032& 1090.1\pp2.3&               18\pp6&&      ---&                ---&&  7.66\pp0.09& 5.81\pp0.27\nl
2.057\pp0.006& 1129.7\pp2.3& 13.4$_{-1.4}^{+6.2}$&&      ---&                ---&& 13.64\pp0.17&   6.6\pp0.6\nl
2.092\pp0.025&    1135\pp13&     68$_{-40}^{+88}$&&      ---&                ---&& 13.46\pp0.04& 9.60\pp0.18\nl
\enddata
\tablenotetext{}{Fit function: see text. All errors correspond to unreduced
$\Delta\chi^2=1$. Fit range, frequency resolution and number of degrees of
freedom for the 1100 and 800\,Hz QPO fits were 200--2000\,Hz, 10\,Hz and 171
for observation 1 orbit 2, 512--1536\,Hz, 5\,Hz and 200 for observation 2 orbit
1 and 200-2000\,Hz, 5\,Hz and 356 for observation 2 orbit 2, respectively.}
\end{deluxetable}

\end{document}